\documentclass[12pt,preprint]{aastex}

\def\kmskpc{{\rm\,km\,s^{-1}{kpc}^{-1}}}

\def\mathnew{\mathsurround=0pt}
\def\simov#1#2{\lower .5pt\vbox{\baselineskip0pt
    \lineskip-.5pt\ialign{$\mathnew#1\hfil##\hfil$\crcr#2\crcr\sim\crcr}}}

\def\'#1{\ifx#1i{\accent"13\i}\else{\accent"13#1}\fi}

\begin{document}    

\shorttitle{The Sun was not born in M~67}  
\shortauthors{Author1 et al. 2010}

\title{The Sun was not born in M~67}

\author{B\'arbara Pichardo, Edmundo Moreno, Christine Allen}

\affil{Instituto de Astronom\'ia, Universidad Nacional
  Aut\'onoma de M\'exico, A.P. 70-264, 04510, M\'exico, D.F.}

\author{Luigi R. Bedin}

\affil{Space Telescope Science Institute, 3700 San Martin Drive,
 Baltimore, MD 21218, USA}

\author{Andrea Bellini}

\affil{Dipartimento di Astronomia, Universit\`a di Padova,
  Vicolo dell'Osservatorio 3, 35122 Padova, Italy}

\author{Luca Pasquini}

\affil{European Southern Observatory, Karl-Schwarzschild-Str. 2,
  85748 Garching bei M\"unchen, Germany.}

\begin{abstract} 
Using the most recent proper-motion determination of the old,
Solar-metallicity, Galactic open cluster M~67, in orbital computations
in a non-axisymmetric model of the Milky Way, including a bar and 3D
spiral arms, we explore the possibility that the Sun once belonged to
this cluster.  We have performed Monte Carlo numerical simulations to
generate the present-day orbital conditions of the Sun and M~67, and
all the parameters in the Galactic model. We compute 3.5 $\times$
10$^5$ pairs of orbits Sun-M~67 looking for close encounters in the
past with a minimum distance approach within the tidal radius of
M~67. In these encounters we find that the relative velocity between
the Sun and M~67 is larger than 20 km/s. If the Sun had been ejected
from M~67 with this high velocity by means of a three-body encounter,
this interaction would destroy an initial circumstellar disk around
the Sun, or disperse its already formed planets. We also find a very
low probability, much less than 10$^{-7}$, that the Sun was ejected
from M~67 by an encounter of this cluster with a giant molecular
cloud. This study also excludes the possibility that the Sun and M~67
were born in the same molecular cloud. Our dynamical results
convincingly demonstrate that M67 could not have been the birth
cluster of our Solar System.

\end{abstract}


\section{Introduction}\label{intro}

In addition to its intrinsic interest for being the birth place of the
only life form we know, the origin of the Solar System provides
important constraints to the current paradigms of star and planet
formation. Identifying some of the history of the Earth's climate (ice
ages, extinctions, etc.) to our place in the Galaxy, knowing some of
the Sun's orbital dynamic history would shed some light to important
matters on different disciplines, not only Astronomy. In this
direction, an interesting question is how the stellar birth
environment influences the formation and evolution of a planetary
system.

Although there is a fraction of stars that was likely formed in
isolated environments, studies from the last decades have revealed
that most stars were born within groups and clusters
\citep{LL95,C00,AM01,LL03,PCA03,AMG07}, and the Sun does not seem to
be the exception. There are several indications that the Solar System
was formed in a gravitationally bound cluster; these include the
observed excitation of the Kuiper belt and the extreme orbital
elements of Sedna \citep{BTR04} and some other bodies in the Kuiper
belt. Moreover, the presence of short-lived radioisotopes in
meteorites indicates that the Solar System formed close to at least
one massive star \citep{WR00,GV00,MC00,HDH04,Ad10}. Several papers
have been devoted to investigate the size of this cluster based on the
observational constraints posed by our planetary disk, such as the
Kuiper belt, which represents a very fragile and restrictive entity.

In a relatively dense star cluster, with a stellar density $\approx$
1000 ${\rm pc}^{-3}$, the Solar System should be able to survive for
as long as 250 Myr before its disruption becomes likely \citep{Ad10},
which can also be inferred from the fact that the inclination angles
of the orbits of Neptune and Uranus are not due to perturbations by
flybys \citep{G95}. Thus, on one hand, considering the long timescales
involved for the Sun to have a mild encounter, it seems necessary that
it was formed in a long-lived cluster --which occurs for about 10\% of
stellar population-- \citep{Ad10}. Only relatively large bound
clusters (of more than a thousand stars), are expected to live that
long \citep{KAH01,LGB05}. Therefore, the Solar System could have been
born in a very large cluster and survive \citep{SC01,MW09}, provided,
for example, that it spent enough time in the outer regions
\citep{Ad10}. On the other hand, an encounter that took place more
than 10 Myr after the Oort cloud started forming, which probably would
have occurred in a dense cluster, would have compromised its formation
\citep{LMD04, Ad10}. In the past several years, observations have
started to place constraints on the plausible distance ranges to
clusters in the Solar neighborhood that could have give birth to the
Sun (those within 2 kpc from the Sun). If our parent cluster was
relatively small, as others in the Solar neighborhood, it probably
contained a few thousand stars, and had a size of a few pc
\citep{AL01,Ad10}. Thus, inferred from the need of a likely nearby
supernova explosion, \citet{Ad10} proposes that the Sun was probably
not too far from the cluster center ($\sim$0.2 pc), where the density
is higher and the more massive stars are placed. In his Table 2,
\citet{Ad10} summarizes some Solar System properties and constraints
on the Solar birth cluster.

The history of the Sun (and its siblings) in the Galaxy, from its
birth to the present has long been a subject of general interest. In
this direction, identifying some of the siblings of the Sun would
provide a direct observational constraint on its birth cluster, with
constraints to the number of stars in the cluster, and even to the IMF
if siblings were found over a range of stellar masses
\citep{PZ09,BPB10,BHKF10}.

In his work, \citet{PZ09} considered the constraints of the Sun's
birth cluster and simulated the orbital evolution of stars in a 1-pc
virial-radius dissolved cluster, with total mass of $\approx$ 10$^3$
M$_{\odot}$, along a Sun-like orbit in an axisymmetric potential
model for the Galaxy. He concluded that, depending on how quickly the
cluster became unbound, between 10\% and 40\% of the Sun's siblings
should be located within 1 kpc of its present location.  \citet{BPB10}
simulated the orbits in the Sun's birth cluster, starting from an
assumed birth place for the Sun, obtained from tracing back the Sun's
orbit over 4600 Myr in a simplified axisymmetric Galactic
potential. They generated a birth cluster and integrated forward in
time in order to find the present day phase space distribution of the
siblings that remain close to the Sun. Using this phase space
distribution, they made a first attempt to identify candidate siblings
of the Sun by searching in the Hipparcos Catalogue, and in the
Geneva-Copenhagen survey of the Solar neighborhood \citep{HNA09}.
They did not find convincing Solar siblings within 100 pc from the
Sun, although they only examined a small fraction of the nearby stars.
 
Making a two dimensional, 2D, numerical experiment like that done by
\citet{PZ09}, but
now using a Galactic potential with stationary and transient spiral
arms, \citet{MA11} show that due to the effect of the spiral arms
there is an appreciable drift of the stars from the original position
of the parent cluster, and only if this cluster had $\approx$
10$^4$ stars there is a good chance of finding Solar siblings within
100 pc from the Sun. Also including a 2D spiral arms potential,
\citet{BBMV11} have found two possible Solar siblings in a sample of
162 candidate stars from the Hipparcos Catalogue, showing that the
spiral arms favor the close encounters of the Sun with these two stars.

Now, an interesting question is: if we come from a stellar cluster, is
part of this cluster still there? Starting from this idea and
considering the very similar metallicities, ages, and distance from
the Galactic center of the Sun and M~67, it becomes tempting to place
the Sun origin within this open cluster.

One of the peculiar and perhaps most intriguing aspects of M~67 is its
chemical composition similar to that of the Sun. All recent high
quality works, based on observations of both, evolved and Solar type
stars, indicate an impressive similarity between the M~67 chemical
composition and the Solar one \citep{TE00,RS06,PP08,FJ10}. The
similarity in composition and age is so close that several authors
have indicated in M~67 stars the best-ever Solar analogues so far
discovered \citep{PB08,OK11,CdNBMD11}, closer than any other field
star in the Solar neighborhood. Such a similarity is particularly
interesting in the context of this work, because it is very tempting
to apply to the Sun the concept of chemical tagging, and to associate
therefore our star to M~67.

Chemical tagging postulates, on the basis of the similarity of
chemical composition observed in open clusters and stellar groups,
that chemical composition can be used to determine the common origin
of stars \citep{FBH02,DFA07,BHKF10}.  This very interesting idea is
supported by the evidence that most stars in open clusters share the
same chemical composition, within the observational uncertainties
\citep{PR04,RS06,DFA07}.  Most importantly, this is not limited to Fe
or metals, but is also valid for neutron capture elements. Similarity
in S and R processes are very important because these elements are
sensitive to the local environment, therefore to the specific location
of the star birth. The cluster chemical homogeneity does not yet
demonstrate that if two stars share the same abundance, they share the
same birthplace. However, by compiling chemical abundances for many
nearby open clusters and comparing them, \citet{DF09} show that large
differences are present in the elemental abundances of these clusters,
strongly indicating that the variability of elements such as Mg, Na,
Zr and Ba, appears quite high even for clusters within $\sim$1 kpc or
less from the Sun. Even if the spread of the data can be influenced by
the compilation of different authors, and by possible systematic
trends between computations of the abundances in dwarfs and giants,
the difference seems large enough to guarantee that clusters with
similar [Fe/H] abundances may well have different chemical
composition, especially as far as neutron capture elements are
concerned. This would suggest that the probability for two stars to
share the abundance pattern should be low if they were not born from
the same cloud.

There is no doubt that, according to this criterion, M~67 is by far
the best known birth-place candidate for the Sun. It has been known
for a long time that the Sun is slightly more metal rich with respect
to the majority of the stars in the Solar neighborhood \citep[][see
  however Casagrande et al. 2011, for a different result]{WG95} and
the hypothesis that the Sun has moved from its birth radius has been
studied in the past \citep{WF96}.

Respecting the age similarity between the Sun and M~67, the age of the
Sun has been determined in several studies \citep
[e.g.,][]{G89,BPW95,GD97,DFet99,BSP02,BBWCH05,MYJ07,HG11}. Its mean
value is around 4.57 Gyr, with a small uncertainty ($\approx$ 1 Myr)
for the formation of the Solar System, but for the Sun itself this
uncertainty being $\approx$ 0.1 Gyr, or even larger, 0.2 Gyr, according
to \citet{DFet99}. The age of M~67 has also been estimated, with several
results: e.g. 4.2 $\pm$ 1.6 Gyr \citep{OK11}; 3.87$^{+0.55}_{-0.66}$
Gyr \citep{CdNBMD11}; 3.9 $\pm$ 0.1 Gyr \citep{Bell10a}; within 3.5
and 4 Gyr \citep{SDK09}; within 3.5 and 4.8 Gyr \citep{YBP08};
4.0 $\pm$ 0.4 Gyr \citep{VS04}; within 3.7 and 4.5 Gyr \citep{RFRI98};
4.0 $\pm$ 0.5 Gyr \citep{DDGP95}; 4.0$^{+1.0}_{-0.5}$ Gyr \citep{DGG92}.
The mean age of M~67 is $\approx$ 4.0 Gyr with a mean uncertainty
$\approx$ 0.5 Gyr. Thus, the Sun's age is approximately within the
2$\sigma$ error bar of M~67's age, and
with their similarity in metallicity, there is some probability, to be
quantified, that the Sun may have born in M~67, or close to this
cluster.

In this work we investigate the reliability of this hypothesis by
computing backward in time the orbits of the Sun and M~67 in a
Milky-Way-like Galactic potential, including the effects of the spiral
arms and a Galactic bar, looking for close encounters. Our study
considers a three dimensional potential for the spiral arms, with the
orbital computation being fully 3D.  We use a recent determination of
the absolute proper motion of M~67 given by \citet{BBP10}, and recent
revisions of the Solar velocity \citep{SBD10} and other Galactic
properties \citep{BCHet05, CHBet09,BRet11}.

This paper is organized as follows. In Section \ref{pm67sol} we
provide data for the Sun and M~67 used to compute their Galactic orbits.
The Galactic mass distribution used in our computations is presented in
Section \ref{modelo}. The employed numerical scheme is given in Section 
\ref{numer}. We present and discuss our results in Section
\ref{resultados}. Our conclusions are given in Section \ref{concl}.

\section{M~67-Sun parameters}\label{pm67sol}

M~67 (NGC 2682) is a well studied, nearby ($\sim$900 pc) cluster. Its
proximity has made it an ideal target for a wide range of studies from
radial velocities \citep{MLG86} and proper motions
\citep{S77,GGL89,BBP10}, to late-type stellar evolution, binary
fractions \citep{MMJ93} and dynamical structure \citep{DS10}. Its
resemblance to the Sun in age and metallicity makes this cluster one
of the key objects in our Galaxy.

In a recent paper, \citet{BBP10} made use of the state-of-the-art
ground-based reduction techniques for ground-based wide-field imagers
\citep{ABP06,YBP08,BPB09} to determine for the first time the absolute
proper motion of M~67 using faint background galaxies as reference. In
this work we decided to adopt their determination, since it is a pure
differential measurement and does not rely, as other works do, on
complex registrations to the ICRS system through a global network of
objects. From \citet{BBP10}, we list in Table \ref{tab.pm67sol} the
absolute proper motion of M~67, and also its radial velocity, v$_r$,
and distance, r, with respect to the Sun, along with their
corresponding uncertainties.

In our computations, four parameters needed for the Sun are its
velocity components with respect to the local standard of rest,
$(U,V,W)_{\odot}$, and its position perpendicular to the Galactic
plane, z$_{\odot}$.  The Solar velocity $(U,V,W)_{\odot}$ has recently
been revised by \citet{SBD10}, with a significant increase in the
value of $V_{\odot}$, in the direction of Galactic rotation. This
Solar velocity is listed in Table \ref{tab.pm67sol}, taking the $U$
component negative towards the Galactic center. The listed
uncertainties are as considered by \citet{BRet11}, who use the new
Solar velocity to obtain a revised value for the ratio of ${\Theta}_0$
and $R_0$, the circular rotation speed and Galactocentric distance of
the local standard of rest. This ratio and a weighted average of
$R_0$, also given by \citet{BRet11}, lead to a corresponding value of
${\Theta}_0$. These values of $R_0$ and ${\Theta}_0$ given by
\citet{BRet11} are also employed in our computations, and are
considered in the following Section. They are within the wide
intervals obtained by \citet{MB10} from an analysis of motions of a
sample of masers in star-forming regions.

The z$_{\odot}$ position is discussed by \citet{R06}, giving several
recent estimates that have been obtained by different authors. From
his Table 1 we have taken the representative z$_{\odot}$-interval as
20$\pm$5 pc. With a given initial (i.e. present) z$_{\odot}$ position
of the Sun in the considered Galactic potential, we add z$_{\odot}$ to
the initial z position of M~67, obtained with the usual convention
that the Sun is on the Galactic plane. The z$_{\odot}$ interval is
also listed in Table \ref{tab.pm67sol}.

\section{The Galactic model}\label{modelo}

We have employed a detailed Milky-Way semi-analytic model to compute
the orbits of the Sun and M~67. This model has axisymmetric and
non-axisymmetric components, the latter due to spiral arms and a
Galactic bar. To build the Galactic model we start with the 
axisymmetric Galactic model of \citet{AS91}, which consists of three
components: disk, spherical bulge, and spherical halo. All the mass
in the spherical bulge is now employed to build the Galactic bar, and
a small fraction, discussed below, of the total mass of the disk is
employed to build the spiral arms. Thus, the only axisymmetric 
components in the final model are the diminished disk and the
spherical halo. 

The inclusion of the spiral arms and the Galactic bar does not modify
the total mass of the original axisymmetric model. Thus, the mean
circular rotation speed on the Galactic plane is maintained; in
particular ${\Theta}_0$ $\approx$ 220 km/s at a Galactocentric
distance $R_0$ = 8.5 kpc, which are the values of these parameters in
the \citet{AS91} model. These values of $R_0$ and ${\Theta}_0$ are now
modified in the non-axisymmetric model to those given by
\citet{BRet11}: $R_0$ = 8.3$\pm$0.23 kpc, ${\Theta}_0$ = 239$\pm$7
km/s.  This is done by first scaling the \citet{AS91} model to a pair
of values ($R_0$,${\Theta}_0$) coming from the Monte Carlo sampling in
our numerical scheme (discussed in Section \ref{numer}), thus
modifying the total masses of the original components, following with
the transformation of the final bulge into the Galactic bar, and
taking a fraction of the final disk to build the spiral arms.

In the following two subsections we comment on some parameters of the
Galactic spiral arms and the Galactic bar. Table \ref{tab.param}
lists the values of these parameters.

\subsection{The Galactic Spiral Arms}\label{brazos}

\subsubsection{Geometry}\label{brazos.geom}

From the Spitzer/$GLIMPSE$ database, \citet{BCHet05} and
\citet{CHBet09} have given new results on the structure of the
Galactic spiral arms, as well as the Galactic bars in the inner region
of our Galaxy. As pointed out by \citet{CHBet09}, although the
geometry of the Galactic spiral structure is the most problematic, it
appears that our Galaxy has two grand-design spiral arms: the
Scutum-Centaurus and Perseus arms, associated with overdensities in
the old stellar disk, and two additional secondary arms, the
Sagittarius and Norma arms, associated with gas
overdensities. \citet{SCWH10} discuss the difference between these two
types of arms, suggesting, as in \citet{MHY04}, that the gaseous
spiral arms are the response of the gas to the main stellar arms.

In our model we include only a spiral pattern that represents the two
main stellar arms.  Our three dimensional (3D) orbital computations
require a 3D potential for these arms. We adopted the spiral arm model
given by \citet{PMM03}, which consists of a 3D superposition of
inhomogeneous oblate spheroids along a given spiral locus, adjustable
to better represent the available observations of the Galactic spiral
arms.
 
\begin{figure}
\plotone{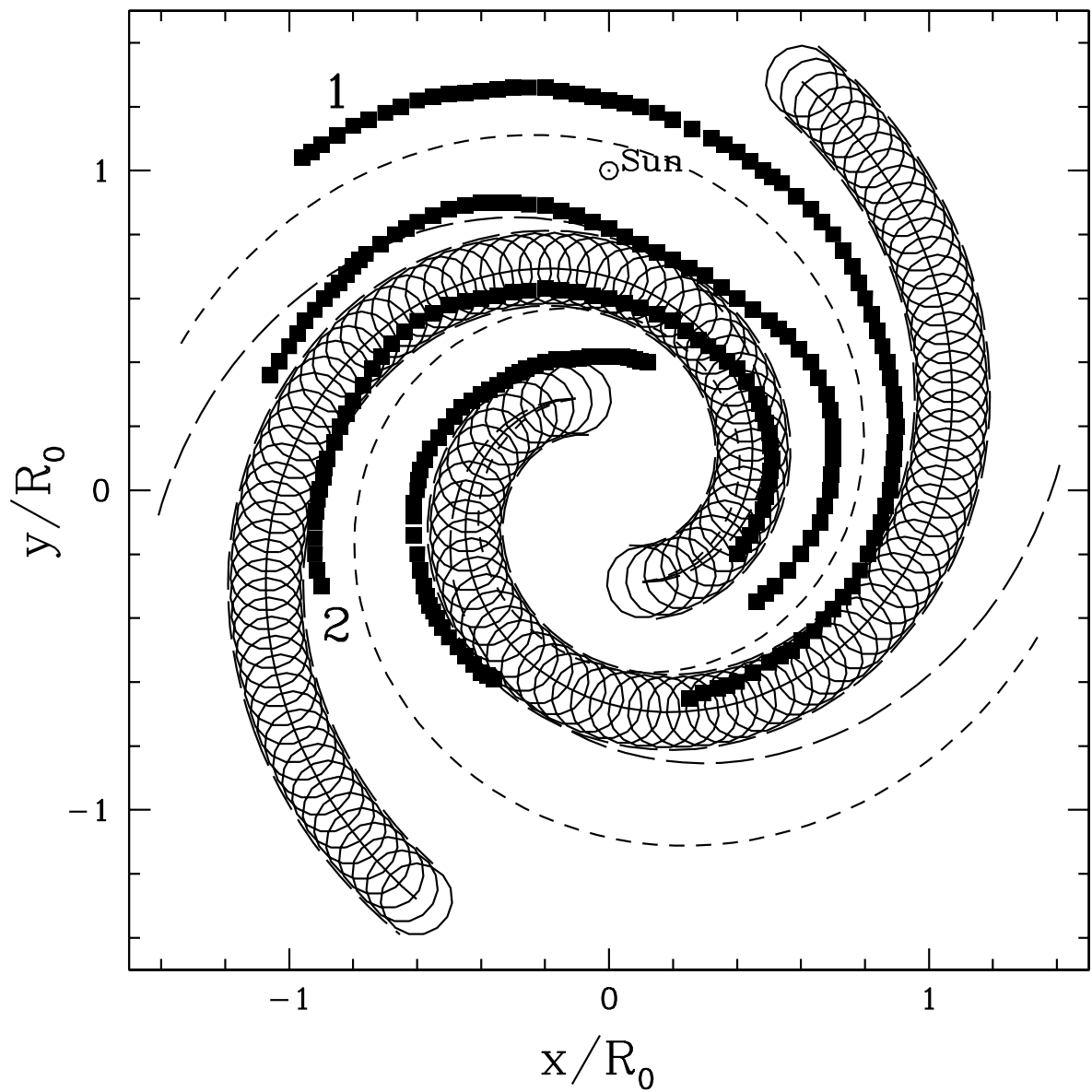}
\caption {Reproduction of figure 2 in \citet{D00}, illustrating in our
model the superposition of spheroids (circles) along the stellar spiral
locus (continuous lines, with pitch angle 15.5$^\circ$) in the K band,
obtained by \citet{D00}.
The short-dashed lines mark a pitch angle limit of 12$^\circ$, and
the long-dashed lines a 19$^\circ$ limit.}
\label{fig1}
\end{figure}

In Figure \ref{fig1} we give details of the geometry of the spiral
arms in our model. This distance-scaled figure on the Galactic plane
is a reproduction of figure 2 in \citet{D00}. The spiral traces shown
with black squares are the observed gaseous spiral arms. The two
continuous lines, on which we illustrate the superposition of
spheroids (shown as circles), are the two stellar spiral arms obtained
by \citet{D00} in the K band, with a pitch angle of 15.5$^\circ$.
These stellar arms are not those stellar arms obtained in the
Spitzer/$GLIMPSE$ data; the latter correspond to the arms marked with
1 (Perseus) and 2 (Scutum-Centaurus) in Figure \ref{fig1} (thus, there
are two gaseous arms that approximately coincide with the two main
stellar arms). However, due to the uncertain geometry of the spiral
structure \citep{CHBet09}, we have considered an interval of values
for the pitch angle of our two spiral arms. In Figure \ref{fig1} the
short-dashed lines mark a lower limit of 12$^\circ$ in the pitch
angle, and the long-dashed lines an upper limit of 19$^\circ$, as in
the second solution obtained by \citet{D00} (but with a different
spiral locus). Thus, in our numerical simulations we take the pitch
angle of the spiral arms as $i$ = 15.5$\pm$3.5$^\circ$; the two
Spitzer/$GLIMPSE$ stellar arms are approximately contained in this
pitch-angle range.

The radial-extent interval of the spiral arms is taken as [2.6,12]
kpc, with $R_0$ = 8.5 kpc. This interval is scaled with the value of
$R_0$ given by the Monte Carlo sampling.

\subsubsection{Strength}\label{brazos.fuerza}

The strength of the spiral arms is related with their total mass,
which is a small fraction of the disk's mass. The distribution of this
mass along the arms is taken as exponential, falling with the radial
scale length of the exponential disk modeled by \citet{BCHet05}:
$H_{\star}$ = 3.9$\pm$0.6 kpc, using $R_0$ = 8.5 kpc (although the
\citet{MN75} disk in the \citet{AS91} model has not strictly an
exponential density fall).

To quantify the strength of the arms we compute the function $Q_T$
\citep{ST80,CS81},
which is the ratio of the maximum azimuthal force of the spiral arms
at a given Galactocentric distance on the Galactic plane, to the radial
axisymmetric force at that distance. The maximum value of $Q_T$
over the radial extent of the spiral arms, called
$Q_s$ = $(Q_T)_{\rm max}$, is a measure of the strength of the spiral
arms. This parameter has been computed
by \citet{BLS04} and \citet{BVSL05} in a sample of spiral galaxies. 

\begin{figure}
\plotone{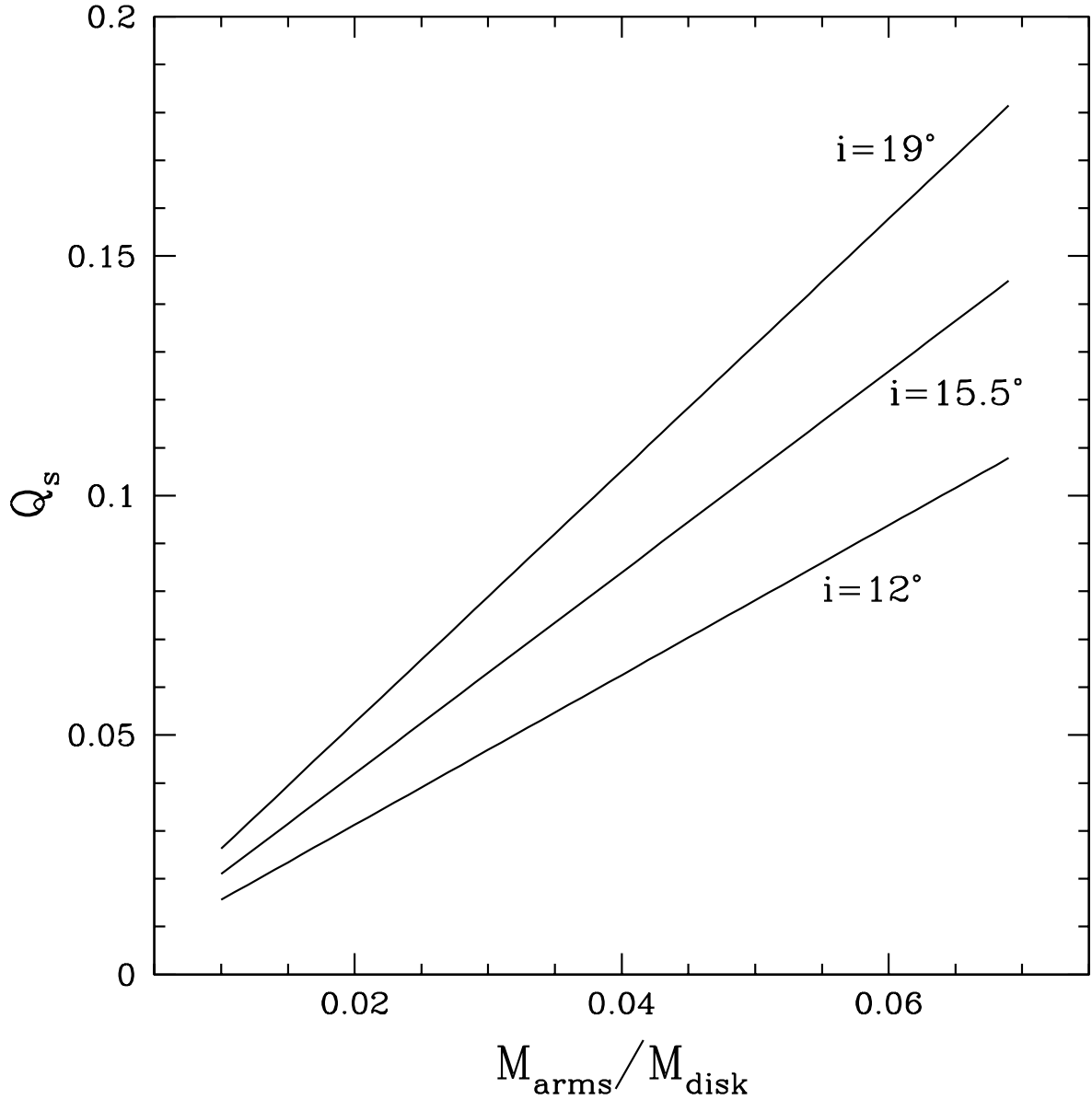}
\caption {The parameter $Q_s$, which gives a measure of the strength
of the spiral arms, as a function of the ratio
$M_{\rm arms}/M_{\rm disk}$ for three values of the pitch angle $i$.
In this figure we take $R_0$ = 8.3 kpc and ${\Theta}_0$ = 239 km/s.}
\label{fig2}
\end{figure}

To exemplify in our model the typical values expected for $Q_s$, 
we take the central values $R_0$ = 8.3 kpc, ${\Theta}_0$ = 239 km/s
of the ($R_0$,${\Theta}_0$) distributions, and
find $Q_s$ for an interval in the ratio $M_{\rm arms}/M_{\rm disk}$.
Figure \ref{fig2} shows the results for three values of the pitch
angle $i$.

In their sample of 147 spiral galaxies, which is dominated by Sbc and
Sc galaxies, \citet{BVSL05} find that 75\% of them have $Q_s$ $\leq$
0.20 . For Sbc galaxies, as our Galaxy, $Q_s$ is approximately less
than 0.25 . Thus the intervals in $M_{\rm arms}/M_{\rm disk}$ and $i$
considered in Figure \ref{fig2} are appropriate for the Galactic
spiral arms in our model. Specifically, we take
$M_{\rm arms}/M_{\rm disk}$ = 0.04$\pm$0.01;
the Gaussian sampling considered in our computations
(Section \ref{numer}) will extend this range, and also the range in
$i$, allowing for acceptable values of $Q_s$ in Figure \ref{fig2}. 
With this ratio $M_{\rm arms}/M_{\rm disk}$ and the 1$\sigma$
variations in $R_0$, ${\Theta}_0$, the mass in the stellar spiral
arms lies in the interval 2.7 -- 5.4 $\times$ 10$^9$ M$_{\odot}$.

\subsubsection{Angular velocity}\label{brazos.omega}

\citet{G11} has given a recent review of different methods to
determine the angular velocity of the Galactic spiral arms,
${\Omega}_S$, as well as those for the angular velocity of the Galactic
bar. Some results are: ${\Omega}_S$ = 24 -- 26 $\kmskpc$ finding the
birthplaces of open clusters \citep{DL05}; ${\Omega}_S$ = 20, 40
$\kmskpc$ with gas flow models \citep{BEG03,MHY04}; ${\Omega}_S$ =
20 -- 30 $\kmskpc$ with kinematics of OB and Cepheids stars
\citep{Fet01}; ${\Omega}_S$ $\approx$ 18 $\kmskpc$ finding two
families of closed orbits in the reference frame of the spiral arms,
which could be associated with the Pleiades/Hyades and Coma Berenices
moving groups \citep{QM05}; ${\Omega}_S$ = 18, 25 $\kmskpc$ from
kinematic fits to five moving groups in the Solar neighborhood
\citep{CH07}. To represent all these results, in our numerical 
simulations we take
${\Omega}_S$ = 24$\pm$6 $\kmskpc$, as listed in Table \ref{tab.param}.

\subsection{The Galactic Bar}\label{barra}

\subsubsection{Geometry and strength}\label{barra.geomf}

For the Galactic bar, we consider the prolate potential given by
\citet{PMM04}, which approximates a model of \citet{F98} of COBE/DIRBE
observations of the Galactic center. In our computations the specific
3D shape of the bar, prolate or triaxial, is not very important,
because the orbits of the Sun and of M~67 will lie far from the region 
of the bar. A prolate model is therefore convenient to facilitate the
computations (also, the COBE/DIRBE Galactic bar approximates a prolate
figure). In Table \ref{tab.param} we list the bar's semi-major axis,
scale lengths, and axial ratio; the scale length of the prolate bar
along its minor axis is taken as the mean of those along the two minor
axes of the triaxial bar models of \citet{F98}. All the lengths of the
bar will change under the Monte Carlo sampling of $R_0$. The 1$\sigma$
variations in $R_0$, ${\Theta}_0$ give a total mass of the bar of 1.5
-- 1.8 $\times$ 10$^{10}$ M$_{\odot}$, which lies in the range 1 -- 2
$\times$ 10$^{10}$ M$_{\odot}$ of estimated values
\citep[e.g.,][]{K92,Z94,DAH95,B95,SUet97,WS99}.

\subsubsection{Orientation and angular velocity}\label{barra.fiomega}

The present orientation of the bar's major axis has been determined
in several studies \citep[e.g.,][]{BGet91,WS99,BG02,BG05,MNQ07}. The
average of the mean values of the angle between the bar's major axis
and the Sun-Galactic center line is around $\phi$ $\approx$ 25$^\circ$.
\citet{G02} suggests $\phi$ = 20$^\circ$ as a good working value.
We take this angle for the present-day orientation of the Galactic bar.

There is also a long list of studies to estimate the bar's angular
velocity, ${\Omega}_B$, \citep[and references therein]{G11}.
\citet{G11} concludes from his review that the most likely range in
${\Omega}_B$ is 50 -- 60 $\kmskpc$. Thus, in our computations we take
${\Omega}_B$ = 55$\pm$5 $\kmskpc$, as listed in Table \ref{tab.param}.

\section{The numerical scheme}\label{numer}

The parameters listed in Tables \ref{tab.pm67sol} and \ref{tab.param},
with their corresponding uncertainties, are employed in the numerical
simulations. The listed uncertainties are considered as 1$\sigma$
variations, and a Gaussian Monte Carlo sampling generates the
parameters to compute the present-day positions and velocities of the 
Sun and M~67, along with the needed parameters in the Galactic
potential. There 
are 14 variables to be sampled, 8 from Table \ref{tab.pm67sol}, and 6 
from Table \ref{tab.param}. For a sampled $R_0$, the lengths associated
with the bar were transformed, and the scale length of the spiral
arms, $H_{\star}$, was taken as the corresponding scaled value of the
mean $H_{\star}$ = 3.9 kpc with $R_0$ = 8.5 kpc. We used the Gaussian
random-number-generator routine \texttt{gasdev} given by
\citet{PTV92}. For each generated set of parameters the orbits of the
Sun and M~67 were simultaneously computed backward in time, in the
time-varying Galactic potential, up to the Solar age, 4.57 Gyr,
allowing for a 0.2 Gyr uncertainty. The high-precision Bulirsch-Stoer
algorithm implemented by \citet{PTV92} was used in all the orbital
computations.

We ran $N_{\rm tot}$ = 3.5 $\times$ 10$^5$ pairs of orbits Sun-M~67,
searching for close encounters in the past; especially in the assigned
time interval for the age of the Sun, $({\Delta}t)_{\rm Sun}$ =
$-$4.57$\pm$0.2 Gyr. However, close encounters outside this interval
were also considered relevant, since, as commented in Section
\ref{sibl}, the Solar System could live within M~67 during 2 or 3 Gyr
before the Sun could escape from this cluster, thus occurring a close
encounter Sun-M~67 in a backward-time computation.

Two major assumptions in our computations are that the structure of
the spiral arms and the bar, as well as their angular velocities,
remain unchanged during the 4.77 Gyr covered in each run. There are
some studies which suggest that the Galactic bar may be a long-lived,
few Gyr old structure \citep{NBCB96,S99,CW02}. This appears not to be
the case for spiral arms in galaxies, more likely being transient
structures \citep{FBSMK11,FRDLW11,S11}. In fact, as shown in the next
Section, the spiral arms can produce a significant effect on the
orbits of the Sun and M~67, and our results suggest that the initial
hypothesis of a common origin of the Sun and M~67 might be more
sustainable taking the strength of the spiral arms as time dependent;
thus modeling these arms as transient features.

\section{Results and discussion}\label{resultados}

\subsection{The Sun as a sibling of M~67's stars ?}\label{sibl}

If the Sun is a sibling of M~67's stars, a convenient criterion for
close encounters Sun-M~67 along their orbits through the Galaxy, is
that the minimum approach distance, $d_{\rm min}$, is less than the
tidal radius of the cluster; i.e., at the time of the encounter the Sun 
is within the region of bounded stellar motions of M~67's stars.
Recently, \citet{HPAT05} have computed N-body models for M~67. Their
best model 2, which reproduces many properties of the present state of
M~67, has an initial tidal radius of 31.8 pc.  During the cluster's
evolution this tidal radius decreases, due to the interaction with the
Galactic potential. Thus, to consider the early phases in the
cluster's evolution, we analyze close encounters for which 
$d_{\rm min}$ $\leq$ 20 pc and $d_{\rm min}$ $\leq$ 30 pc.

\begin{figure}
\plotone{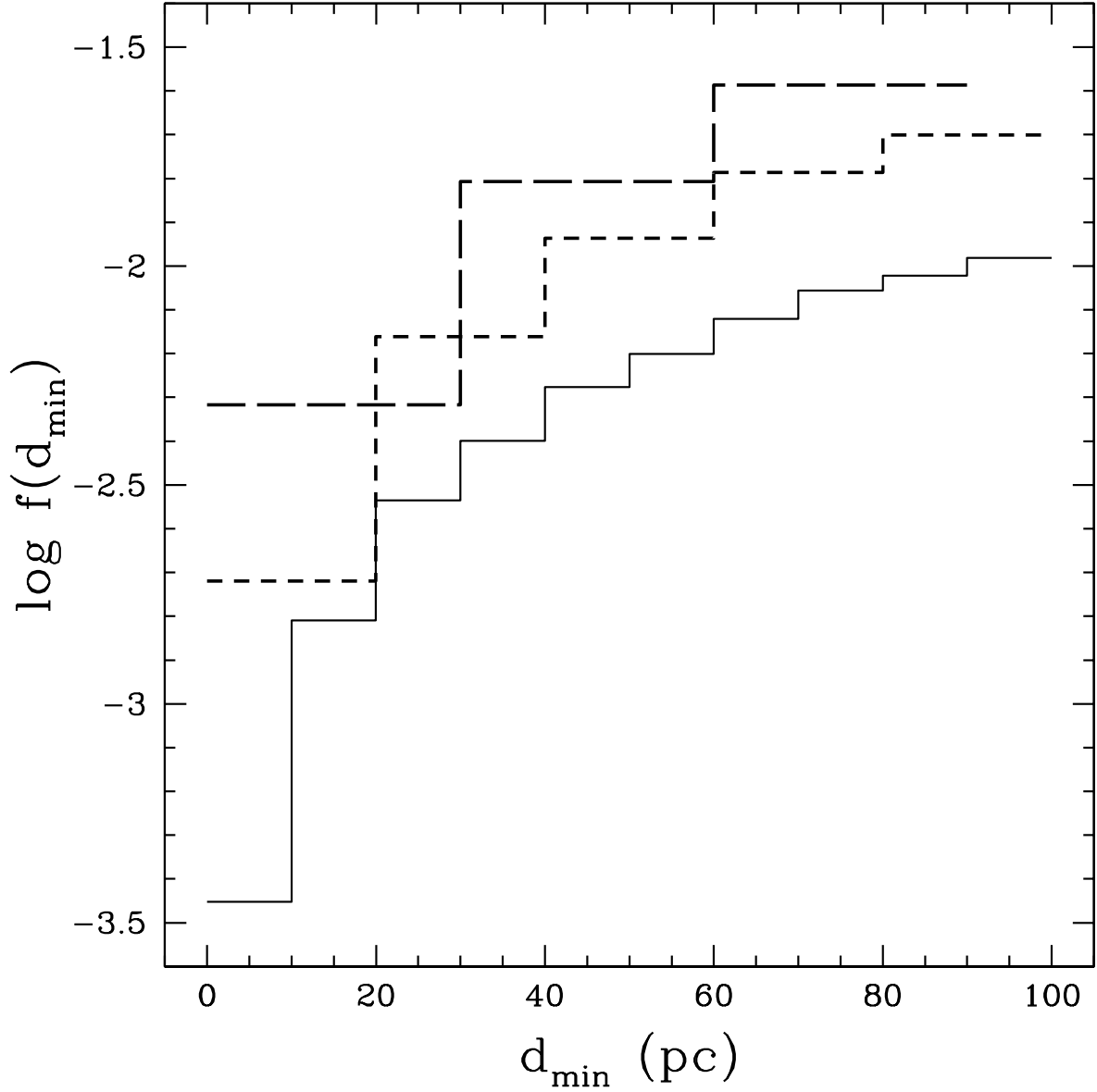}
\caption {Logarithm of the close-encounters distribution
f($d_{\rm min}$), for $d_{\rm min}$ bins of 10 pc (continuous line),
20 pc (short-dashed line), and 30 pc (long-dashed line).
The encounters occur at any time, up to the Sun's age.}
\label{fig3}
\end{figure}

To illustrate the form of the resulting distribution of close
encounters in our computations, up to $d_{\rm min}$ = 100 pc, Figure
\ref{fig3} shows the logarithm of the distribution of close encounters
Sun-M~67, f($d_{\rm min}$) = (number of encounters in the $d_{\rm min}$
interval)/$N_{\rm tot}$, $N_{\rm tot}$ = 3.5 $\times$ 10$^5$. This 
figure includes all the encounters in a given $d_{\rm min}$ interval,
independently of the time they occurred. The continuous line is for
10-pc bins in $d_{\rm min}$, the short-dashed line for 20-pc bins,
and the long-dashed line for 30-pc bins. As a start, there are many
encounters in the intervals of interest, $d_{\rm min}$ $\leq$ 20 pc,
$d_{\rm min}$ $\leq$ 30 pc.

\begin{figure}
\plotone{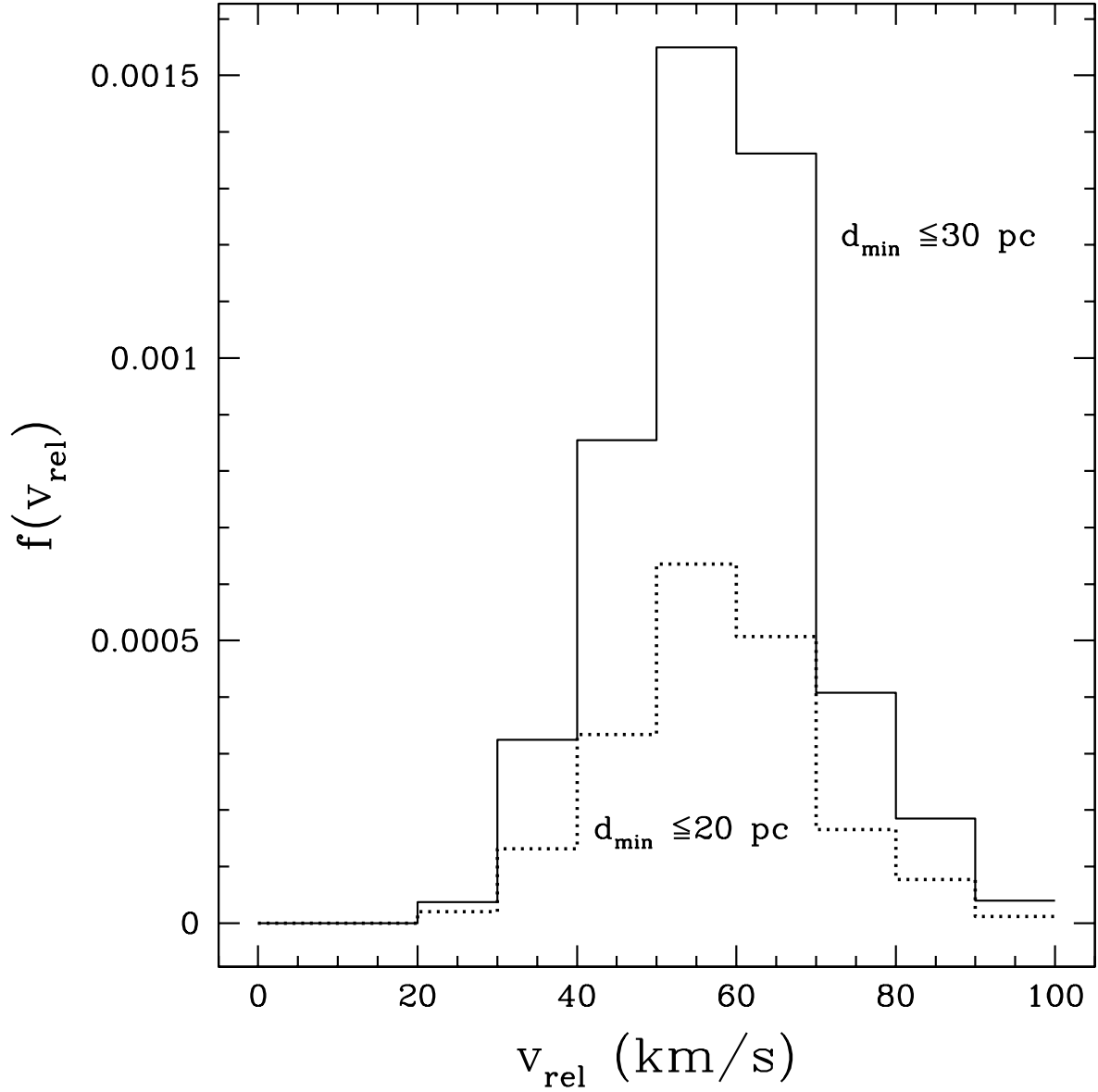}
\caption {The distributions in relative velocity between the Sun and
M~67 for the close encounters in Figure \ref{fig3} with
$d_{\rm min}$ $\leq$ 20 pc and $d_{\rm min}$ $\leq$ 30 pc.}
\label{fig4}
\end{figure}

Taking from Figure \ref{fig3} only the close encounters with
$d_{\rm min}$ $\leq$ 20 pc, and $d_{\rm min}$ $\leq$ 30 pc, we show in
Figure \ref{fig4} their distribution in relative velocity between the
Sun and M~67 at the time of the encounter, f($v_{\rm rel}$) = (number of
encounters in the $v_{\rm rel}$ interval, with $d_{\rm min}$ $\leq$ 20
pc or $d_{\rm min}$ $\leq$ 30 pc)/$N_{\rm tot}$. The peaks in the
distributions are around 50 -- 60 km/s, and no encounters were obtained
with $v_{\rm rel}$ $\leq$ 20 km/s.

The same analysis can be made especially for those close encounters
which occurred within the assumed time interval for the Sun's age,
$({\Delta}t)_{\rm Sun}$ = $-$4.57$\pm$0.2 Gyr. Figure \ref{fig5} shows 
their distributions in relative velocity, with $d_{\rm min}$ $\leq$ 20
pc, and $d_{\rm min}$ $\leq$ 30 pc.
The majority of these encounters have relative
velocities larger than 30 km/s, and as already shown in Figure
\ref{fig4}, there are no encounters with $v_{\rm rel}$ $\leq$ 20 km/s.

With the assumption that the Sun was born in a star cluster,
\citet{Ad10} estimates that the planetary system, formed in a time
scale about 10 Myr, can survive within the cluster for $\sim$ 0.25 Gyr
if the cluster's star density is $n_{\star}$ $\sim$ 10$^3$ pc$^{-3}$.
This surviving time increases if $n_{\star}$ decreases. The M~67
N-body model of \citet{HPAT05} has an initial $n_{\star}$ = 50
pc$^{-3}$ within the half-mass radius, i.e. M~67 was not an initial
dense system. With this $n_{\star}$, the surviving time for the Solar
System within M~67 could be 2 or 3 Gyr, followed by the escape of the
Sun. Thus, considering close encounters outside the Sun's-age time
interval is relevant in this respect. However, the important
constraint of the survival of the planetary system imposes a
corresponding constraint on the relative velocity between the Sun and
M~67 at the time of the Sun's ejection (encounter, in our
backward-time computation).

\subsubsection{The Sun ejected from M~67 by a three-body encounter ?}
\label{tres}

The limit $v_{\rm rel}$ $>$ 20 km/s obtained in our computations
results a high value compared to the smooth escape-velocity ejection
from a star cluster, with a typical internal velocity dispersion of
$\sim$ 1 km/s. According to \citet{GGL89}, the present-day velocity
dispersion in M~67 is 0.81$\pm$0.10 km/s. However, in these cluster
environments, three-body encounters are typical processes capable to
produce high-velocity ejections; particularly a binary-single star
encounter. This type of interactions has been amply studied
\citep{HB83,H83a,H83b,H84,H85,H93, HHM96}. A binary star in the
stellar cluster can capture another star in the cluster, a bound
triple system is formed, and after some time one of the stars is
ejected (this process is known as $resonance$ $scattering$;
\citet{HB83}); or the captured star is ejected in the interaction
(this is called a $flyby$; \citet{HB83}).  For high velocity
ejections, hard binaries play the important role.  In this case the
ejected star leaves the three-body system with a velocity of the order
of the internal binary velocities \citep{H93}.  Thus, this ejection
velocity is approximately \citep{Aa06}

\begin{equation} v_{\rm ejec} \simeq \left(\frac{G(m_1+m_2)}{2a}\right)^{1/2},
\label{eq1} \end{equation}

\noindent with $m_1$, $m_2$ the masses of the stars in the binary
system, and $a$ its semi-major axis. Thus, for a hard binary with
equal mass components, and individual masses of order $\sim$ 1
M$_{\odot}$, the ejection velocity is of order $(GM_{\odot}/a)^{1/2}$.
Taking this ejection velocity $v_{\rm ejec}$ $>$ 20 km/s, as obtained
in our numerical simulations, it results approximately $a \leq$ 2 AU
(the M~67 N-body model of \citet{HPAT05} has an upper cut-off at $a$ =
50 AU for the initial binaries in the system). \citet{HI85} show that
in an encounter of a hard binary with a single star, the typical
minimum close distance between any two of the three stars can be less
than the original semi-major axis $a$ of the binary. In a $resonance$
$scattering$ this minimum distance can be of the order 0.01$a$.  Thus,
if the Sun were to be captured by a hard binary with $\approx$ $a
\leq$ 2 AU, as obtained from our results, this small-distance
interaction would destroy an initial circumstellar disk around the
Sun, and no planets would be formed; or the planets would be dispersed
if they are already formed.  \citet{Ad10} comments that since the
radius of the planetary system is $\approx$ 30 AU (the position of
Neptune), and a close encounter truncates an existing initial disk at
a distance $\approx$ 1/3 of its impact parameter, the required
minimum distance in a close encounter suffered by the Solar System is
$\approx$ 90 AU; much larger than the $\leq$ 2 AU in the triple-system
interaction. Even with an order of magnitude increase in the total
mass of the binary star, the semi-major axis, $a \leq$ 20 AU, is still
below the 90 AU limit.  The situation would be worse if the ejected
Sun from the triple system comes from a $resonance$ $scattering$,
where the strong interaction between the three bodies may last orders
of magnitud longer than the initial binary period \citep{H93}, and
there are frequent close approaches.

Thus, with the obtained results, we conclude that the Sun was not
ejected from M~67 by a three-body encounter. Although we do find close
encounters Sun-M~67 with an appropriate $d_{\rm min}$, the relative
velocity of the encounter is too high.

\subsubsection{The Sun ejected from M~67 by an encounter with a giant
molecular cloud ?}\label{dos}

A high star-ejection velocity from a star cluster can also be produced
by an encounter with a giant molecular cloud
\citep[and references therein]{GPZBALSL06}. \citet{SSS79} find that the
mass of these clouds, $M_n$, is in the range 10$^5$ -- 3 $\times$
10$^6$ $M_{\odot}$, and assuming their virial equilibrium 
\citet{SRBY87} give
a mass-radius relation $M_n$ = 540$R_n^2$ $M_{\odot}$, with the
clouds's radius $R_n$ in parsecs.

Analytic estimations of the change in velocity of a star in the star
cluster, at a distance $R_{\star}$ from its center, due to the
encounter with a giant molecular cloud, can be obtained in the cases
of distant and head-on encounters. For a distant encounter, in the
impulse approximation, the change in velocity in a direction
perpendicular to the relative velocity cloud-cluster, is \citep{S87}

\begin{equation} {\Delta}v = \frac{2GM_nR_{\star}}{b^2V},
\label{eq2} \end{equation}

\noindent with $b$ the impact parameter, and $V$ the magnitude of the
relative velocity cloud-cluster. With Plummer models for both cloud
and cluster, with median radii $a_1$, $a_2$, this relation applies
with $b \geq$ 5max($a_1$, $a_2$) \citep{GPZBALSL06}.

For a head-on encounter, $b$ = 0, and using Plummer models, the change
in velocity in a direction perpendicular to the relative velocity
cloud-cluster, is \citep{GPZBALSL06}

\begin{equation} {\Delta}v = \frac{2GM_nR_{\star}}{V(R_{\star}^2+
a_n^2)},
\label{eq3} \end{equation}

\noindent with $V$ as in Eq. (\ref{eq2}), and $a_n$ = $R_n$/2. 

Taking the upper limit $M_n$ $\approx$ 3 $\times$ 10$^6$ $M_{\odot}$
for the mass of the giant molecular cloud, the corresponding radius
obtained with
$M_n$ = 540$R_n^2$ $M_{\odot}$ \citep{SRBY87} is
$R_n$ $\approx$ 75 pc. M~67 has an initial tidal radius of 31.8 pc,
according to the model of \citet{HPAT05}. Thus, Eq. (\ref{eq2})
applies with $b \geq$ 5$a_n$ = 5$R_n$/2 $\approx$ 190 pc. In our
numerical simulations, at all close encounters Sun-M~67, which occur
near the Galactic plane, M~67 has a velocity $\approx$ 20 -- 40 km/s
in the direction perpendicular to the Galactic plane (i.e., its
$W$-velocity component), and this is the dominant component of its
peculiar velocity. On the other hand, the typical random velocity
of clouds is $\approx$ 7 km/s \citep{BT08}. Thus, for $V$ in Eqs.
(\ref{eq2}) and (\ref{eq3}) we take $V \geq$ 15 km/s.

With $b$ = 190 pc, $V$ = 15 km/s, and $R_{\star}$ = 30 pc (i.e., a
star, the Sun, at a distance from the center of M~67 equal to M~67's
tidal radius), Eq. (\ref{eq2}) gives ${\Delta}v$ $\leq$ 1.4 km/s,
and Eq. (\ref{eq3}), ${\Delta}v$ $\leq$ 22 km/s. For encounters
cloud-cluster other than distant or head-on, ${\Delta}v$ will lie
between these estimates.

\begin{figure}
\plotone{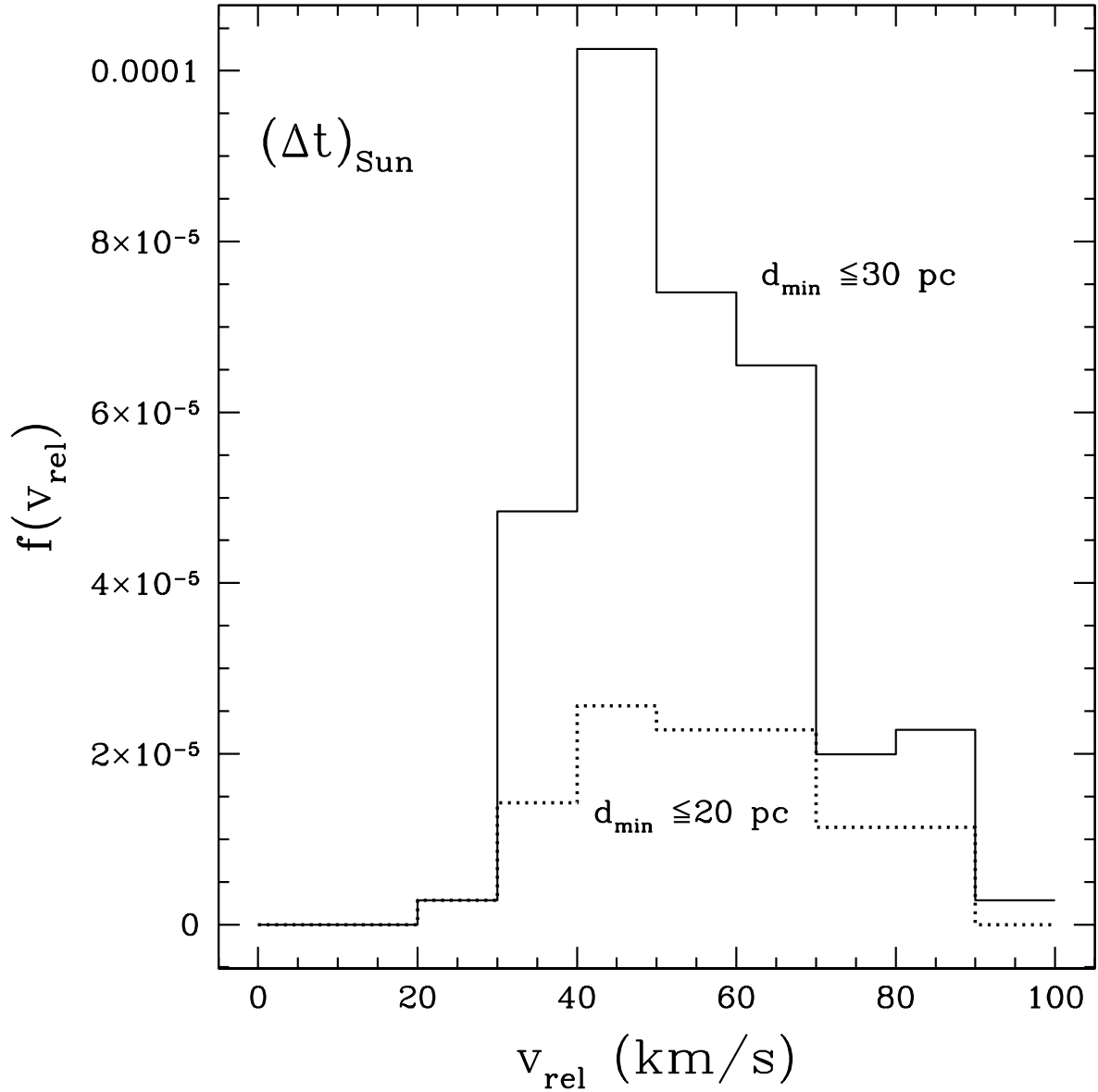}
\caption {The distributions in relative velocity between the Sun and
M~67 for close encounters which occurred in the Sun's-age time
interval $({\Delta}t)_{\rm Sun}$ = $-$4.57$\pm$0.2 Gyr. Encounters with
$d_{\rm min}$ $\leq$ 20 pc and $d_{\rm min}$ $\leq$ 30 pc are
considered.}
\label{fig5}
\end{figure}

Thus, a head-on encounter between M~67 and a most massive giant
molecular cloud, with the Sun being a M~67's star at a distance equal
to M~67's tidal radius, and located appropriately in a direction
perpendicular to the relative velocity between M~67 and the cloud,
could impart to the Sun an ejection velocity of order 20 km/s. These
velocities are obtained in our computations, as shown in Figures
\ref{fig4} and \ref{fig5}, but with a low probability: less than
10$^{-4}$ for encounters Sun-M~67 occurring at any time (Figure
\ref{fig4}), and less than 10$^{-5}$ for encounters Sun-M~67 in the
Sun's-age time interval (Figure \ref{fig5}). The net probability is
even lower, since all the above conditions on the type of encounter
M~67-cloud, mass of the cloud, position of the Sun in M~67, and the
Sun ejected appropriately for its orbit to be of low-z amplitude, must
be fulfilled simultaneously. Furthermore, the $a$ $priori$ probability 
for the Sun being born in a star cluster with appropriate conditions
consistent with present properties of the Solar System is 0.0085
\citep{AL01}. Thus, we conclude that there is a very low probability,
much less than 10$^{-7}$, that the Sun was ejected from M~67 by an
encounter with a giant molecular cloud.

\subsection{The Sun and M~67 born in the same molecular cloud ?}
\label{nube}

\begin{figure}
\plotone{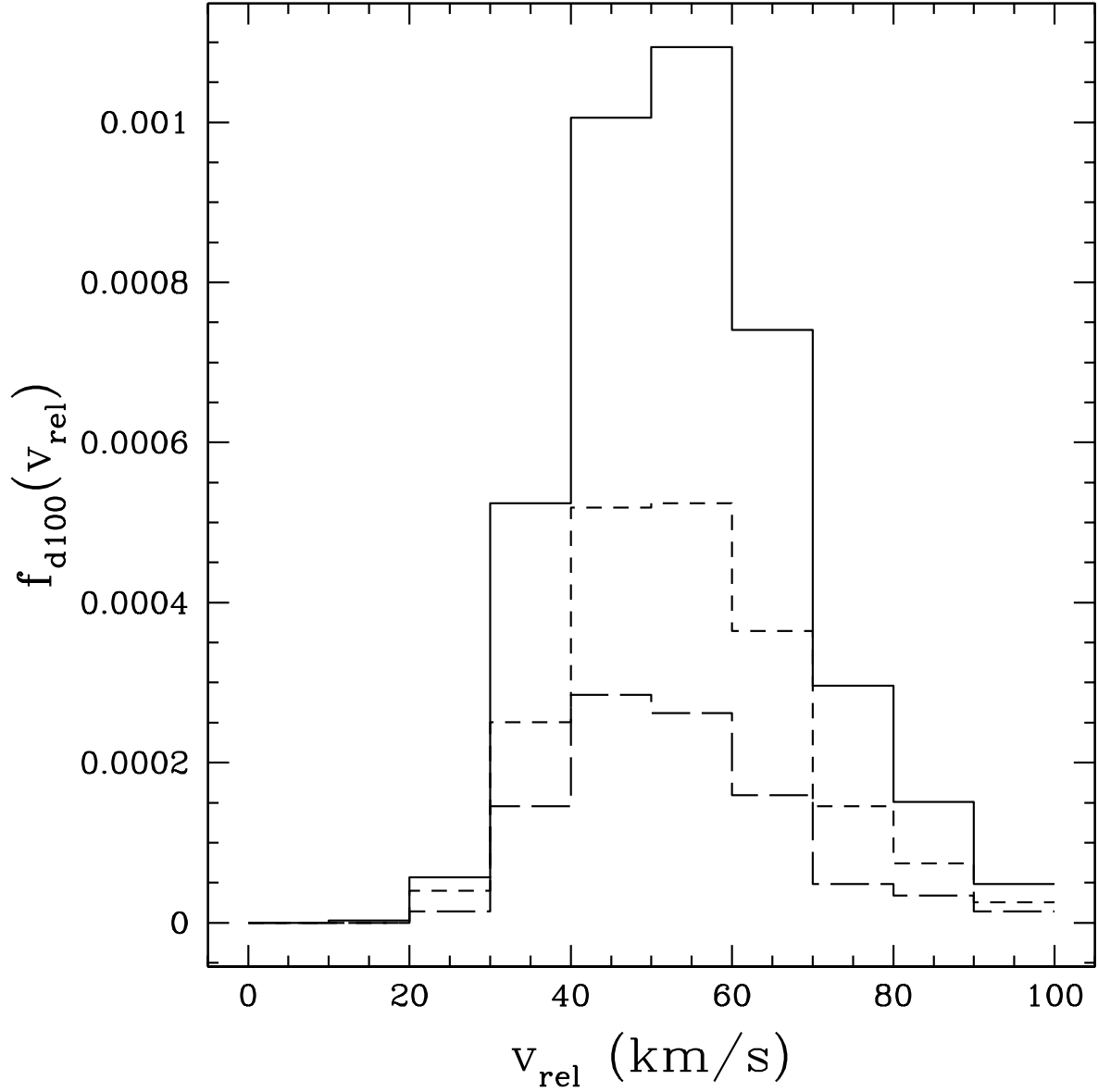}
\caption {Distributions in relative velocity between the Sun and M~67
for close encounters with $d_{\rm min}$ $\leq$ 100 pc, occurring in the
time intervals: $-$4.57$\pm$0.2 Gyr, continuous line;
$-$4.57$\pm$0.1 Gyr, short-dashed line; $-$4.57$\pm$0.05 Gyr,
long-dashed line.}
\label{fig6}
\end{figure}

As long as a molecular cloud has been assembled and no supernovae have
been produced in its interior, the cloud may be approximately
chemically homogeneous \citep{BHKF10}. Thus, another possibility that
can be analyzed with our computations is that the Sun and M~67 were
born in the same molecular cloud.  To test this possibility, we
consider close-encounter distances Sun-M~67 up to 100 pc, say, and
take only such encounters occurring within the time interval for the
Sun's age, or subintervals centered at its mean age.  Figure
\ref{fig6} shows their distributions in relative velocity between the
Sun and M~67, for three time intervals centered at $-$4.57 Gyr:
$-$4.57$\pm$0.2 Gyr, continuous line; $-$4.57$\pm$0.1 Gyr, short-dashed
line; $-$4.57$\pm$0.05 Gyr, long-dashed line. Only in the first
interval there is a close encounter with $v_{\rm rel}$ = 17.8 km/s; 
in all three intervals the rest of encounters have $v_{\rm rel}$ $>$ 20
km/s. This is a high velocity compared with typical random velocities
less than 10 km/s within clouds with sizes less than 100 pc
\citep{L81,SRBY87}. Also, at all these encounters M~67 has a velocity
$\approx$ 20-40 km/s in the direction perpendicular to the Galactic
plane, which also exceeds the typical 7 km/s random velocity between
clouds. Thus, we also exclude the possibility that the Sun and M~67
were born in the same molecular cloud.

\subsection{The effect of the spiral arms}\label{bref}

The conclusions in Sections \ref{sibl} and \ref{nube} depend strongly
on the high relative velocity between the Sun and M~67 at their close
encounters. The relative velocity, $v_{\rm rel}$, between the Sun and
M~67 in a close encounter depends on their respective Galactic
orbits. The high values of $v_{\rm rel}$ obtained in the computations
reflect mainly the different z amplitudes above or below the Galactic
plane reached by their orbits. At the present time, M~67 is above the
Galactic plane, at z $\approx$ 400 pc; the Sun is at z $\approx$ 20
pc. If we compute their orbits in an axisymmetric potential (both will
be box type), the Sun would stay within $|$z$|$ $\approx$ 50 pc, and
M~67 within $|$z$|$ $\approx$ 400 pc. Any close encounter obtained in
this case will have a high $v_{\rm rel}$, since M~67 will have a high
z-velocity when it crosses the low-z region where the Sun moves. A
non-axisymmetric Galactic potential is more promising to investigate
the hypothesis of a common origin Sun-M~67, because the Galactic
orbits are perturbed by the non-axisymmetric components, and $v_{\rm
  rel}$ can reach lower values.  For our problem, the spiral arms play
a key role.  \citet{AMP08} have investigated the effect of the spiral
arms, of the type we are using in the present study, on the orbits of
some globular clusters. They used $M_{\rm arms}/M_{\rm disk}$ = 0.03,
and found a slight difference in z amplitudes compared with the orbits
computed in an axisymmetric potential; the main difference appeared in
the radial motions, parallel to the Galactic plane. However, in our
present study we are extending the range of possible values of $M_{\rm
  arms}/M_{\rm disk}$; values as high as 0.06 -- 0.07 are being
sampled by the Monte Carlo scheme, consistent with a $Q_s$ for a
galaxy with the same Hubble type as the Milky Way (Figure \ref{fig2}).

\begin{figure}
\plotone{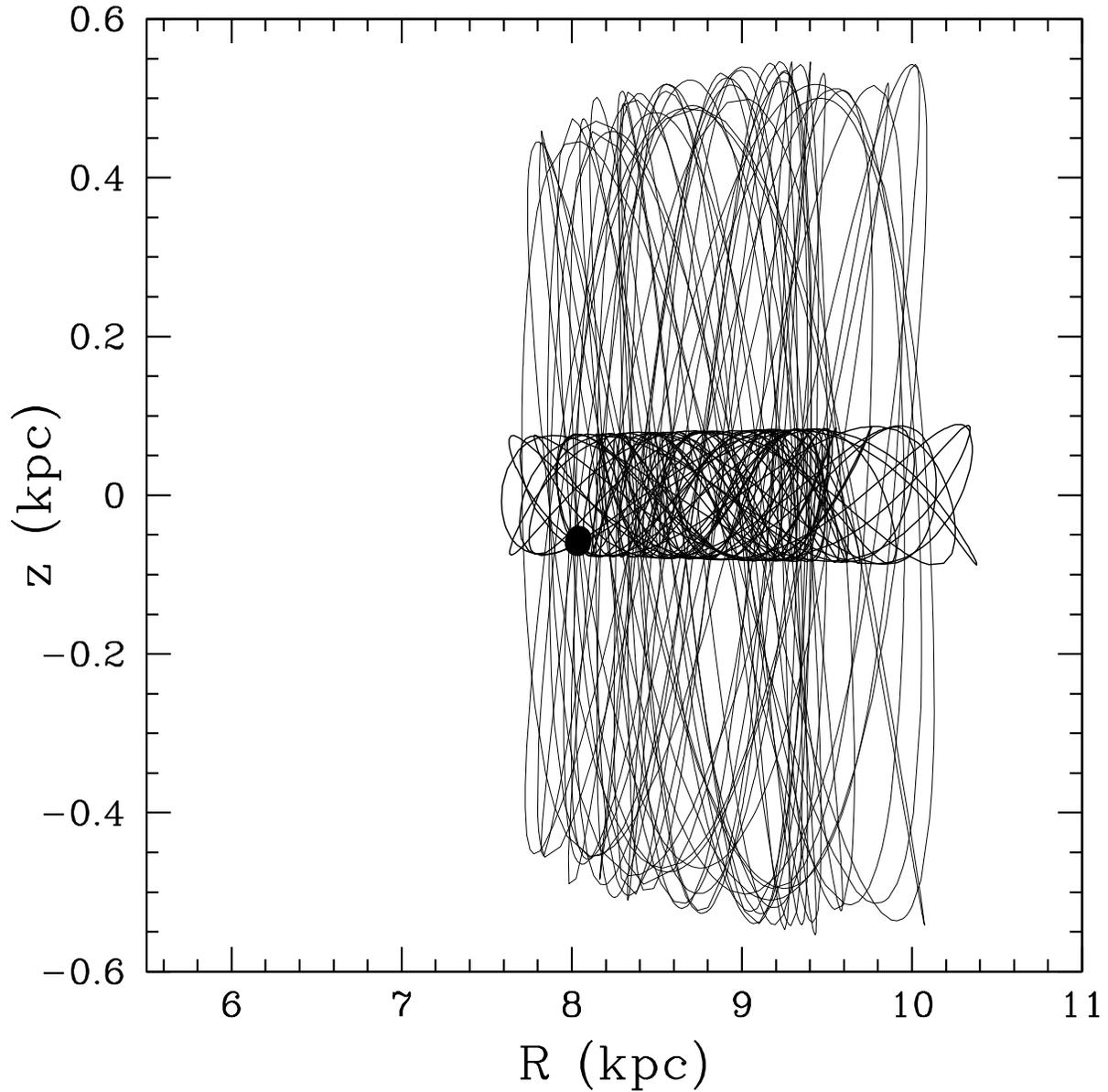}
\caption {Meridional orbits of the Sun (low-z amplitude) and M~67
(high-z amplitude) for a run with a low strength of the spiral arms,
$M_{\rm arms}/M_{\rm disk}$ = 0.0123. The black dot shows the position
of the close encounter.}
\label{fig7}
\end{figure}

Figure \ref{fig7} shows the meridional orbits of the Sun (low-z
amplitude) and M~67 (high-z amplitude) for one of our runs. This
case has $M_{\rm arms}/M_{\rm disk}$ = 0.0123; i.e., a low strength of 
the spiral arms. The black dot shows the position of the close
encounter. Note that there is a slight perturbation on both orbits.
In this case the relative velocity at the encounter is
$v_{\rm rel}$ = 49.6 km/s; a high value.

\begin{figure}
\plotone{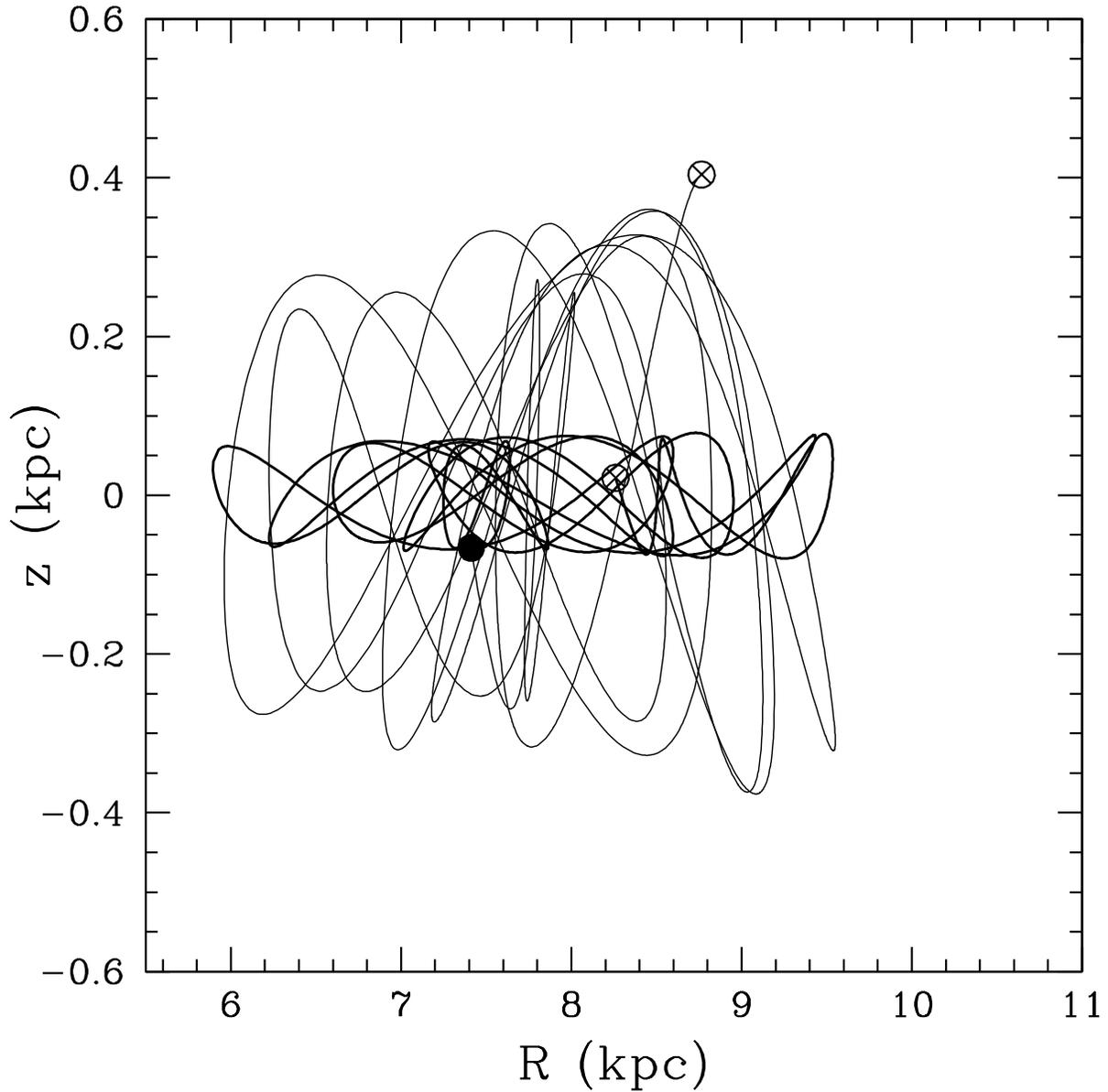}
\caption {As in Figure \ref{fig7}, with the same (R,z) region, here
for a run with a high strength of the spiral arms,
$M_{\rm arms}/M_{\rm disk}$ = 0.0636 . The points $\otimes$ mark the
present-day positions of the Sun and M~67, and the black dot shows the
position of the close encounter.}
\label{fig8}
\end{figure}

With the same (R,z) region as in Figure \ref{fig7}, we show in
Figure \ref{fig8} another case, but now $M_{\rm arms}/M_{\rm disk}$
= 0.0636 .
This is a case with high strength of the spiral arms. The present-day
positions of the Sun and M~67 are marked with $\otimes$, 
and the black dot shows the corresponding position of the
close encounter. Here, the orbit of the Sun shows
a strong perturbation in the radial direction, and M~67's orbit has
also strong perturbations in both radial and z directions. Note
especially that the z amplitude, $|$z$|$$_{\rm max}$, in the orbit of
M~67 around the time of the close encounter, is nearly half of that
reached in Figure \ref{fig7}. In this case the relative velocity at
the encounter is $v_{\rm rel}$ = 24.3 km/s, one of the lowest values of
$v_{\rm rel}$ obtained in our computations.

If it were possible to find situations in which the z amplitude,
$|$z$|$$_{\rm max}$, of M~67's orbit around the time of the close
encounter, were similar to the z amplitude of the Sun, then $v_{\rm
  rel}$ would decrease, and more favorable conditions might be found
to support a common origin of the Sun and M~67. In this respect, an
important question is: how has M~67 acquired its present, high-z
position? One possible answer is that it keeps memory of its birth at
such high z. \citet{VPGFMC10} analyzed a sample of open clusters from
the open cluster catalogue by \citet{DAML02}. The majority of the open
clusters in the sample have low $|$z$|$$_{\rm max}$: 90$\%$ have
$|$z$|$$_{\rm max}$ $<$ 350 pc. For clusters with higher $|$z$|$$_{\rm
  max}$, \citet{VPGFMC10} mention the proposed formation mechanisms at
high z of \citet{MAF99} and \citet{FMFM08}, and also suggest others.
From the WEBDA Open Cluster Database \citep{PM08} \citet{FMFM08} give
two examples of young clusters that are likely born at high-z distance
from the Galactic plane. \citet{F95} also discusses high-z distance
formation scenarios for the old open clusters, like M~67. Thus, M~67
could have been born at a $|$z$|$ distance comparable with its present
z $\approx$ 400 pc. But in this case, if the Sun were to be ejected
from M~67, a high $v_{\rm rel}$ would be needed to put it into its
low-z amplitude orbit, and we obtain the conclusions in Sections
\ref{sibl} and \ref{nube}.

Another possible answer for the present, high-z, position of M~67 is
that it was born in a low-z amplitude orbit, and has been dispersed in
the z direction by the force of the spiral arms, and collisions with
interstellar clouds. M~67 has been able to withstand the tidal force of
the clouds without being disrupted, as theory predicts \citep{S58}.
As we have said above, this situation might be more favorable to a
common origin of the Sun and M~67; the Sun leaving M~67 with a lower
value of $v_{\rm rel}$ when M~67 is in its low-z amplitude phase.

\begin{figure}
\plotone{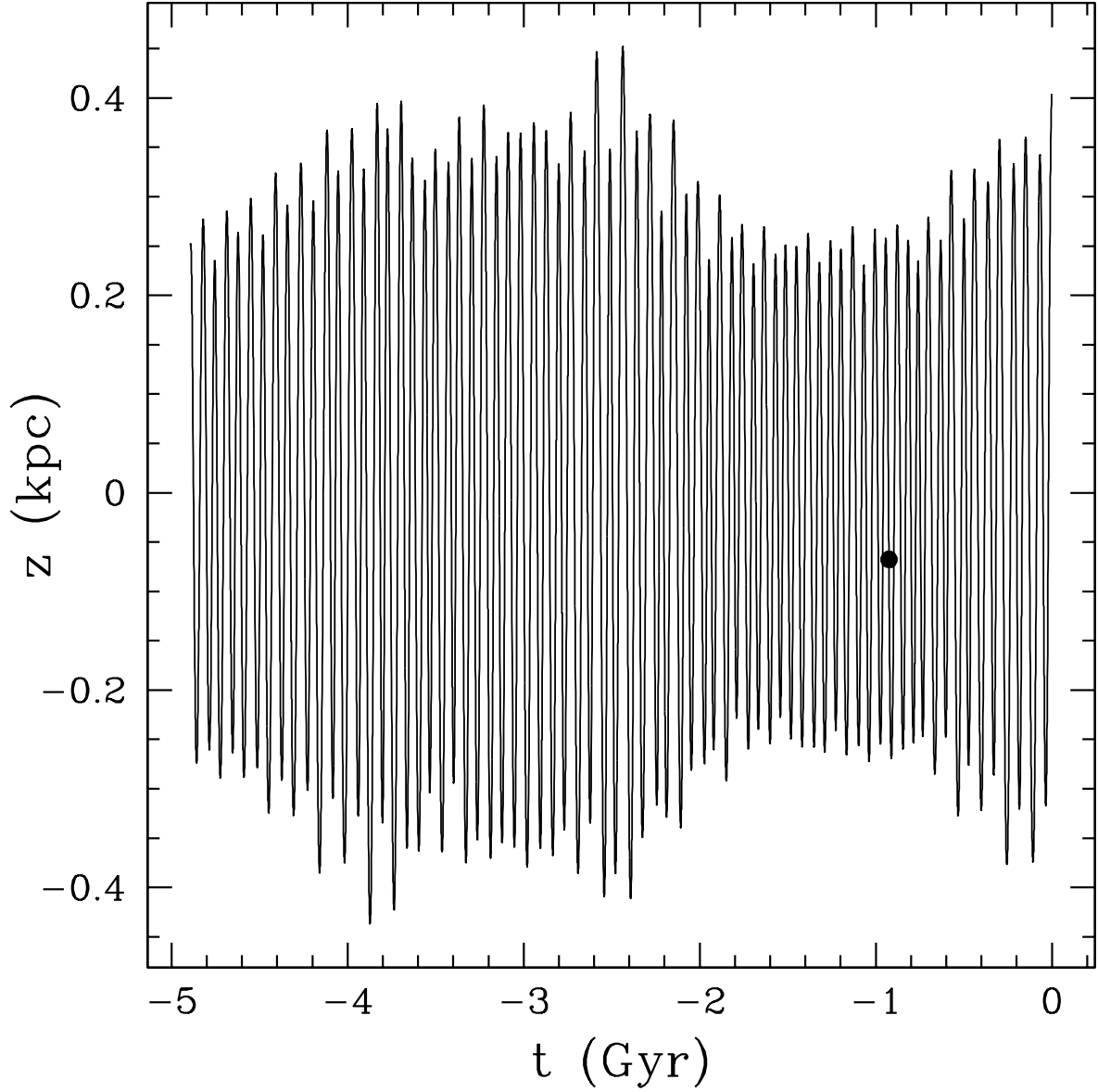}
\caption {Evolution of the coordinate z in the orbit of M~67, for the
case shown in Figure \ref{fig8}, extending the time up to $-$5 Gyr. 
The black dot shows (t,z) at the close encounter Sun-M~67.}
\label{fig9}
\end{figure}

The example presented in Figure \ref{fig8} shows that there is an
increase in the z amplitude of M~67's orbit from the time of the
encounter to the present time; i.e., the orbit is dispersed to high z
in this interval of time. Figure \ref{fig9} shows the evolution of the
coordinate z as a function of time in M~67's orbit for the case in
Figure \ref{fig8}, but the time extended up to 5 Gyr in the past, to
show more clearly the effect of the spiral arms. The black dot shows
the coordinates (t,z) of the close encounter Sun-M~67, which occurred
around 1 Gyr in the past. There is approximately a periodic
increase-decrease in this z coordinate. The example in this
Figure \ref{fig8} and cases with similar behavior in the z coordinate,
suggest that maybe considering non steady spiral arms, their
increase-decrease periodicity in the z coordinate could be changed,
obtaining situations in which this coordinate further decreases. 
This transient-spiral-arms modeling would be more
in line with the conclusions of \citet{FBSMK11}, \citet{FRDLW11}, and
\citet{S11}. This will be done in a future study.

\section{Conclusions} \label{concl}

Considering the resemblance in age, metallicity, and distance from the
Galactic center, of the old open cluster M~67 and the Sun, we explored
in this paper the possibility that the Sun was once a member of
M~67. We employed for this purpose the most recent proper-motion
determination of M~67 in orbital computations in a 3D non-axisymmetric
Milky Way model.  The non-axisymmetric Galactic model includes a bar
and 3D spiral arms, rotating with their particular angular
velocities. The employed parameters to specify the properties of the
bar and spiral arms have been taken from several papers from the
literature. We used a Gaussian Monte Carlo sampling to generate
different present-day orbital conditions for the Sun and M~67, and
parameters of the Galactic model. We ran 3.5 $\times$ 10$^5$
simultaneously-computed pairs of orbits Sun-M~67, looking for close
encounters in the past. The analysis of these close encounters shows
that the corresponding relative velocity between the Sun and M~67 is
larger than 20 km/s. This velocity is too high; a three-body encounter
within M~67, with the Sun being one of the three bodies, and giving
this ejection velocity to the Sun, would destroy an initial
circumstellar disk around the Sun, or disperse its already formed
planets. Thus, the Sun was not ejected by a three-body encounter in
M~67. Also, by analyzing a possible encounter of M~67 with a giant
molecular cloud, we find a very low probability, much less than
10$^{-7}$, that the Sun was ejected from M~67 by such an
encounter. The high values of the relative velocity also exclude the
possibility that the Sun and M~67 were born in the same molecular
cloud. We have illustrated the effect of the spiral arms on the
Galactic orbits of the Sun and M~67. Modeling the spiral arms as
transient features might prove to be compatible with the Sun-M~67
common-origin hypothesis.

\acknowledgments We thank the anonymous referee for a very careful  
review and several excellent suggestions that greatly improved this
work. B.P. and E.M. thank support by grant PAPIIT
IN110711-2. A.B. acknowledges the support by MIUR under program
PRIN2007 (prot.\ 20075TPK9).

\clearpage
\begin{deluxetable}{ccc}
\tablecolumns{3}
\tablewidth{0pt}
\tablecaption{M~67-Sun parameters}
\tablehead{\colhead{Parameter} &\colhead{Value} & \colhead{References}}
\startdata
\cutinhead{M~67}
$({\mu}_{\alpha}\cos{\delta})_{2000.0}$   & $-$9.6$\pm$1.1 mas yr$^{-1}$ & 1 \\
$({\mu}_{\delta})_{2000.0}$   & $-$3.7$\pm$0.8 mas yr$^{-1}$ &  1 \\
v$_r$                         & 33.78$\pm$0.18 km s$^{-1}$   &  1 \\
r                             & 815$\pm$81.5 pc                &  1 \\
\cutinhead{Sun}
$(U,V,W)_{\odot}$  & ($-11.1\pm$1.2,12.24$\pm$2.1,7.25$\pm$0.6)
 km s$^{-1}$  &  2 \\
z$_{\odot}$   &   20$\pm$5 pc  &  ~3 
\label{tab.pm67sol}
\enddata
\tablerefs{
            1)~\citet{BBP10}.
            2)~\citet{SBD10}.
            3)~\citet{R06}.}

\end{deluxetable}

\clearpage
\begin{deluxetable}{lcr}
\tablecolumns{3}
\tablewidth{0pt}
\tablecaption{Sun's Galactocentric distance, local circular rotation
speed, and parameters of the non-axisymmetric Galactic components}
\tablehead{\colhead{Parameter} &\colhead{Value} & \colhead{References}}
\startdata
$R_0$                                   & 8.3$\pm$0.23 kpc     & 1 \\
${\Theta}_0$                            & 239$\pm$7 km/s       & 1 \\
\cutinhead{Spiral Arms}
pitch angle ($i$)                   & 15.5$\pm$3.5$^{\circ}$ & 2 \\
scale length ($H_{\star}$)          & 3.9$\pm$0.6 kpc  ($R_0$ = 8.5 kpc)        & 3 \\
$M_{\rm arms}/M_{\rm disk}$                 &  0.04$\pm$0.01         &   \\
mass                                & 2.7 -- 5.4 $\times$ 10$^9$ M$_{\odot}$           &  \\
pattern speed ($\Omega_S$)        & 24$\pm$6 $\kmskpc$     & 4 \\
\cutinhead{Prolate Bar}
semi-major axis                     & 3.13 kpc  ($R_0$ = 8.5 kpc)            & 5 \\
scale lengths                       & 1.7, 0.54 kpc ($R_0$ = 8.5 kpc)       & 5 \\
axial ratio                         & 0.54/1.7            &   \\
mass                                & 1.5 -- 1.8 $\times$ 10$^{10}$ M$_{\odot}$     &   \\
angle between major axis            &                     &   \\
\,\,\, and the Sun-GC line          & 20$^{\circ}$        & 6 \\
pattern speed ($\Omega_B$)        & 55$\pm$5 $\kmskpc$  & ~ 4
\label{tab.param}
\enddata
\tablerefs{
            1)~\citet{BRet11}.
            2)~\citet{D00}.
            3)~\citet{BCHet05}.
            4)~\citet{G11}.
            5)~\citet{F98}.
            6)~\citet{G02}.}
\end{deluxetable}

\clearpage

\clearpage

\clearpage

\clearpage

\clearpage

\clearpage

\clearpage

\clearpage

\clearpage

\end{document}